\documentclass[twocolumn]{aastex631}
\usepackage{amsmath}
\usepackage{bm}

\begin{document}

\title{On Acceleration of Highest-Energy Cosmic Rays in a Novel Scenario of Magnetar Transients}

\correspondingauthor{Jiro Shimoda}
\email{jshimoda@icrr.u-tokyo.ac.jp}

\author[0000-0003-3383-2279]{Jiro Shimoda}
\affiliation{Institute for Cosmic Ray Research, The University of Tokyo, 
5-1-5 Kashiwanoha, Kashiwa, Chiba 277-8582, Japan}

\author[0000-0001-6010-714X]{Tomoki Wada}
\affiliation{Department of Physics, National Chung Hsing University, Taichung, Taiwan}
\affiliation{Frontier Research Institute for Interdisciplinary Sciences and Astronomical Institute, Graduate School of Science, Tohoku University, Aoba, Sendai, 980-8578, Japan}
\affiliation{Institute for Cosmic Ray Research, The University of Tokyo, 
5-1-5 Kashiwanoha, Kashiwa, Chiba 277-8582, Japan}



\begin{abstract}
Transient phenomena in magnetars have been considered as
possible acceleration sites of ultrahigh-energy cosmic-rays
(CRs), whose energy reaches $\sim200$~EeV,
such as the Amaterasu particle.
However, the process of CR acceleration and the trigger
mechanism of magnetar transients remains unclear.
A recently suggested scenario for the activity predicts
that the magnetar's rotation axis suddenly flips due to
the `Dzhanibekov effect,' resulting in a sudden rise of
the Euler force. The material in the outer layer plastically
flows due to the force and finally fractures in this
scenario. We study the possibilities of ion acceleration
along with this scenario. If the degenerate electrons burst
open from the fractured region like a balloon burst, the pair
plasma formation can be ignited inside the crust. We find that
such pair plasma can emit photons similar to the observed bursts
from magnetars. We also find that the electron stream at the
beginning of the burst phenomenon possibly induces a strong
electric field for a moment, resulting in the acceleration of
$\sim1$~ZeV ion within a timescale of $\sim1$~ps. The nuclear
spallation reactions limit this timescale, and therefore,
high-energy CR `neutrons' from the parenteral nuclei become
proper observational predictions of this scenario: their arrival
time and direction will be correlated with the bursting photon
emissions of the host magnetars. The nuclear spallation of
$\sim$~ZeV nuclei is preferred to explain  $\gtrsim10$~PeV
neutrino events observed by IceCube and KM3Net.
\end{abstract}



\section{Introduction}
\label{sec:intro}
The highest-energy cosmic rays (CRs) with energy above
$10^{20}$~eV$=100$~EeV are one of the most mysterious
particles in the Universe
since the first detection \citep{linsley63}. Recently,
\citet{ta23} detected the second-highest energy CR in
history, $244\pm29{\rm (stat.)}~^{+51}_{-76}{\rm (sys.)}$~EeV,
which is named as `Amaterasu'. The arrival direction of
Amaterasu is highly contentious; Amaterasu seems to come
from a void in the large-scale structure of the Universe,
where energetic astrophysical objects do not exist.
The objects such as radio galaxies hosting relativistic jets,
starburst galaxies, and clusters of galaxies
are usually invoked as the origins of ultrahigh-energy CRs
($>1$~EeV)
\citep[e.g.,][for reviews]{anchordoqui19,globus23}.
\citet{unger24} supported no existence of candidates among
the radio galaxies by cross-matching the astronomical
source catalog and backtrack analysis of Amaterasu. They
pointed out the most straightforward possibility that
Amaterasu was accelerated in an astrophysical transient event
in undistinguished galaxies. Moreover, the
$\sim120^{+110}_{-60}$~PeV muons event, KM3-230213A, recently
discovered by KM3Net Collaboration, implying the arrival of
$35\mathchar`-380$~PeV cosmic neutrino \citep{km3net25}.
This event is thought to imply the existence of CR sources
other than those responsible for lower-energy neutrinos because
the estimated energy flux exceeds the flux in the lower-energy
range observed by IceCube Collaboration \citep{icecube18}.
To explain the above observations, we need to consider the
possibilities of CR acceleration processes in a broad sense,
however this may be challenging.
\par
Magnetars, a type of neutron star with strong magnetic
field (say, $B\gtrsim10^{14}$~G),\footnote{
The original definition is that magnetars are sources
powered by their magnetic energy \citep{duncan92}.}
are one of the candidates for the origins of ultrahigh-energy
CRs \citep[e.g.,][]{arons03,asano06,kotera11}. Some of
the magnetars frequently show bursting activities in
the X-ray/soft gamma-ray band observed as astrophysical
transient events; we refer to such events as `magnetar burst'
in this paper. Although many magnetar bursts are observed and
progressively studied, their details are still unclear
\citep[e.g.,][for reviews]{harding06, hurley08, kaspi17, enoto19}.
The magnetar burst from SGR~1935+2154 (`SGR' is an initialism
of Soft Gamma-ray Repeater and indicates the object name), with
a total energy of $\sim10^{40}$~erg in the X-ray band shows
an association of fast radio bursts (FRBs)
with a total energy of $\sim10^{35}$~erg
\citep{bochenek20,chime20,mereghetti20,li21frb,ridnaia21,tavani21}.
FRBs have raised attention in the past decade
\citep[][for reviews]{zhang20,lyubarsky21,petroff22}.
However, their origins and emission mechanisms are still
under debate.
\par
The SGRs exhibit short, repeating bursts of photons
with a typical energy of $\sim 10$~keV and luminosity of 
$<10^{44}$~erg~s$^{-1}$ \citep{kaspi17}. \citet{hurley08}
mentioned that once an SGR is active, hundreds of bursts
are emitted in a period of minutes. The active periods appear
at random intervals. Outside of these periods, there are no
detectable bursts for years. Therefore, the intrinsic event
rates of the bursts may be unsettled. As a very rare event,
a more intense bursting phenomenon called a giant flare is
observed with a luminosity of $>10^{44}$~erg~s$^{-1}$ and
typical photon energy of $\sim100$~keV \citep{kaspi17}.
The burst from SGR~1935+2154 mentioned above is an exceptional
case, where the photon energy reaches $\sim100$~keV, but the
luminosity is $\sim10^{40}$~erg~s$^{-1}$. We refer to this
exceptional case as simply the FRB-associated burst in this
paper, following \citet{yang21}.
\par
Since our understanding of the magnetar bursts is still
insufficient, a simultaneous study of the ultrahigh-energy CR
acceleration and astrophysical transient events from magnetars
may be preferred, i.e., by the multimessenger astrophysical
methods. In this paper, following a novel scenario of a magnetar
burst proposed by \citet{wada24}, we construct a scenario of
the highest-energy CR acceleration with the energetics of the
bursting activities. Although we make several assumptions and
simplifications, we show that the scenario can be tested by a
combination of the CR experiments, neutrino observations, and
X-ray observations.
\par
The paper is organized as follows: Section~\ref{sec:upset}
reviews the scenario for a magnetar burst proposed by \citet{wada24}
and the burst properties are discussed in section~\ref{sec:burst}.
Consequent ion acceleration in this scenario is discussed
in section~\ref{sec:acc}. Proper observational predictions
of our model are summarized in section~\ref{sec:discussion}.
We use $Q_{,x}=Q/10^x$ in cgs Gauss units unless otherwise noted.

\section{Novel Scenario of Magnetar Burst}
We describe the energetics of a magnetar burst according
to the scenario proposed by \citet{wada24}. Neutron stars
(including the magnetars) sometimes show the so-called
glitch activity, which is interpreted as a sudden change
in the angular velocity \citep[e.g.,][]{fuentes17}, and
its origin is a matter of debates \citep[e.g.,][]{kaspi17}.
\citet{hu24} reported associations between such glitch
activities and an FRB from SGR~1935+2154.
As described below, we consider a situation in which
the angular velocity vector of the magnetar is unstable,
undergoing the so-called Dzhanibekov effect, described by
the classical mechanics of rigid bodies \citep{landau69}.
Although the realization of such a setup has not been
established yet in neutron stars, the connection between
the sudden change of the angular velocity and bursting
phenomena can be studied.
\par
In the following, we define the magnetar outer layer of
its body with the density $\rho\lesssim10^{12}$~g~cm$^{-3}$
and refer the material to the `crust' \citep[e.g.,][]{chamel08}.
The crust is assumed to consist of electrons and iron nuclei
with $(Z,A)=(26,56)$ in this paper.
\par
\subsection{Unstable Free Precession of Magnetar}
\label{sec:upset}
The magnetar with its mass and radius of $M_*\sim1.4~M_\odot$
and $R_*\sim10^6$~cm is assumed to have a triaxial body
with an ellipticity of $\delta$ (not completely spherical
or axisymmetric) and its initial spin axis does not coincide
with any principal axes of inertia. Here, the ellipticity of
$\delta\ll1$ is a free parameter, which can be tested by the
gravitational wave observations, \citep[e.g.,][]{zim80,
fujsawa22,ligo22}, the analysis of pulse profiles 
\citep{shakura98,link07,makishima14,makishima24,kolesnikov22},
and so on. When the angular velocity around the second
principal axis, $\Omega_2$, initially dominates over
the others ($\Omega_2\gg\Omega_1,\Omega_3$, but $\Omega_1\neq0$
or $\Omega_3\neq0$), the direction of the second principal axis
is unstable and eventually `flips' (the Dzhanibekov effect).
\citet{wada24} analyzed the case as an example that the three
eigenvalues of the moment of inertial tensor are $I_3>I_2>I_1$,
where $I_2=(2/5)M_*R_*{}^2$,$I_1=(1-\delta)I_2$, and $I_3=
(1+\delta)I_2$.
\par
Here, we review the Dhzanibekov effect briefly \citep{landau69}.
Under the rigid body approximation, the magnetar body
follows the Euler equations in the body frame $(x_1,x_2,x_3)$ as
%
\begin{eqnarray}
&& \frac{{\rm d}\Omega_1}{{\rm d}t}+\frac{I_3-I_2}{I_1}\Omega_2\Omega_3=0,
\nonumber \\
&& \frac{{\rm d}\Omega_2}{{\rm d}t}+\frac{I_1-I_3}{I_2}\Omega_3\Omega_1=0,
\nonumber \\
&& \frac{{\rm d}\Omega_3}{{\rm d}t}+\frac{I_2-I_1}{I_3}\Omega_1\Omega_2=0,
\label{eq:euler eq}
\end{eqnarray}
%
with the conservation laws of the energy 
%
\begin{eqnarray}
\frac{J_1{}^2}{I_1} +
\frac{J_2{}^2}{I_2} +
\frac{J_3{}^2}{I_3} = 2E,
\label{eq:energy body}
\end{eqnarray}
%
and the angular momentum
%
\begin{eqnarray}
J_1{}^2 +
J_2{}^2 +
J_3{}^2 = J^2,
\label{eq:am body}
\end{eqnarray}
%
where $J_i=I_i\Omega_i$ (i=1,2,3) is the component of
the angular momentum vector, $J=|\bm{J}|$ is the total
angular momentum (given constant), and $E=(1/2)\sum_{i}
I_i\Omega_i{}^2$ is the total energy (given constant),
respectively. In the state-space of $\bm{J}$, the energy
equation~(\ref{eq:energy body}) represents the ellipsoid
with the semi-axes lengths of $\sqrt{2EI_1}$, $\sqrt{2EI_2}$,
and $\sqrt{2EI_3}$, while the angular momentum
equation~(\ref{eq:am body}) represents the sphere with a
radius of $J$. The components of vector $\bm{J}$ varies
according to the Euler equations~(\ref{eq:euler eq})
in the body frame, and the head of $\bm{J}$ moves by pointing
along the line of intersections between the surfaces of
the ellipsoid and the sphere.
\par
\begin{figure*}
\center
\includegraphics[width=15 cm]{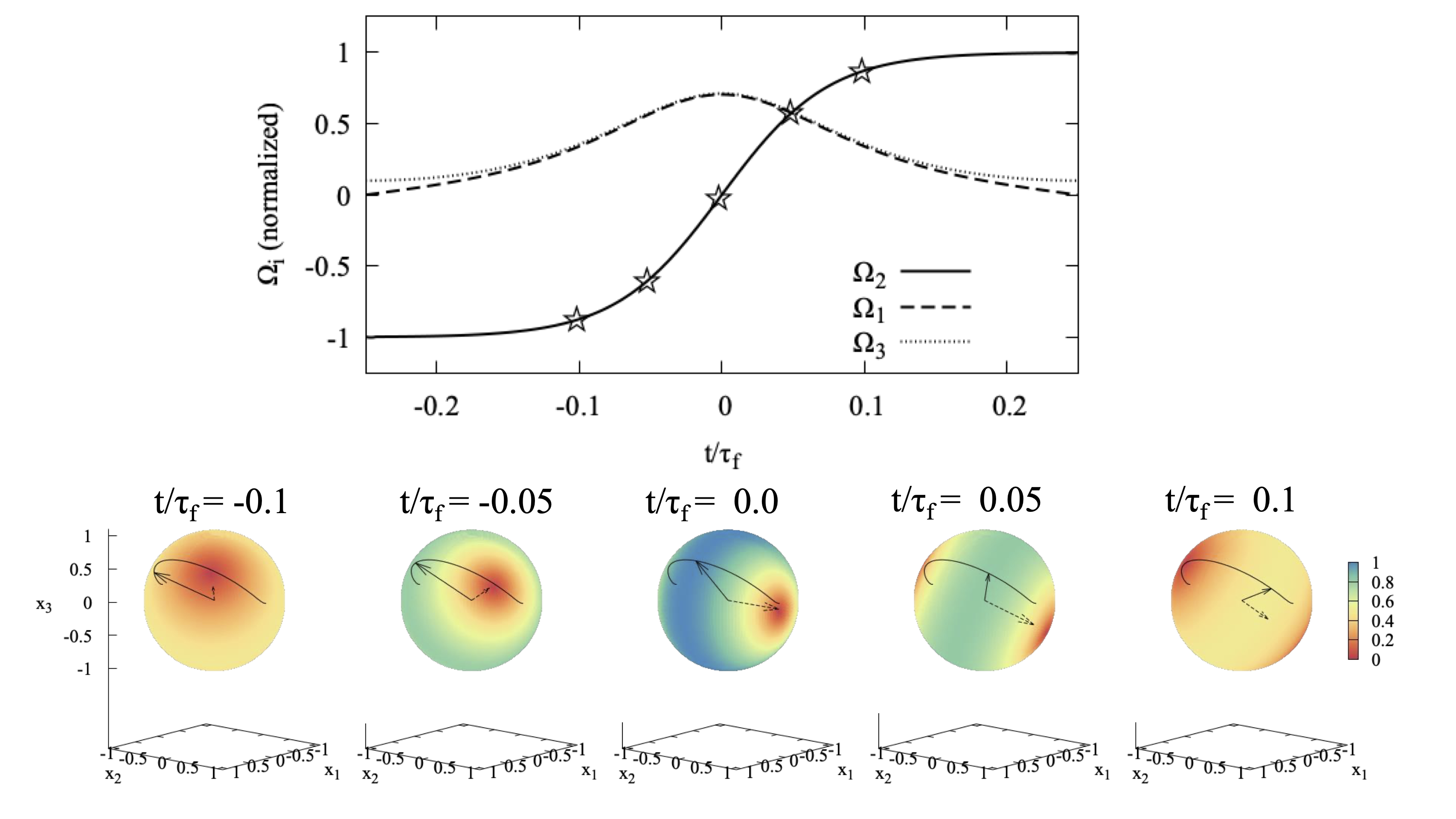}
\caption{
Temporal variation of  $\bm{\Omega}(t)$ in the
body frame. We set normalized parameters as
$I_3=1.0+\delta$,
$I_2=1.0$,
$I_1=1.0-\delta$,
$\delta=10^{-6}$, and the angular velocity vector
of $\bm{\Omega}=(0,-\sqrt{1-0.1^2},0.1)$, where
$\Omega_2=0$ at $t=0$~\citep{landau69}.
The upper panel shows the temporal evolution
of $\Omega_1$, $\Omega_2$, and $\Omega_3$, around
the flip term. The star symbols in the upper panel
indicate the five distinct time points as displayed
in the surrounding panels. At each distinct time,
the Euler force magnitude on the body is shown in
the bottom panels. The color of each bottom panel
indicates the magnitude of the Euler force normalized
by its maximum during the flip.
The  solid arrow shows $\bm{\Omega}(t)      $,
the dashed arrow shows $\bm{\dot{\Omega}}(t)$
(pointing to the red region), and the solid curve
shows the trajectory of the head of
$\bm{\Omega}(t)$, respectively.} 
\label{fig:body}
\end{figure*}

The existence of the intersection line requires the
condition of $2EI_1<J^2<2EI_3$ in our case ($I_3>I_2>I_1$).
When $J^2\rightarrow2EI_1$ ($J^2\rightarrow2EI_3$), the
line becomes approximately a small closed circle, which
asymptotically converses to a pole of the ellipsoid on
the $J_1$ ($J_3$) axis. This case can be regarded as a
stable precession: The direction of $\bm{\Omega}$ does
not significantly change in the body frame. In our case
with $\Omega_2\gg\Omega_1,\Omega_3$ and $I_3\simeq I_2
\simeq I_1$ ($\delta\ll1$), the closed line becomes large:
The direction of $\bm{\Omega}$ drastically changes in
the body frame. This can be regarded as an unstable
situation, called the Dzhanibekov effect. Figure~\ref{fig:body}
shows examples of the temporal variation of the $\bm{\Omega}$
in the body frame. The head of $\bm{\Omega}$ initially along
$\approx\bm{J}/I_2$ follows the line of intersections,
and this motion results in a ``flip" of the body in the
laboratory frame (see also \citet{wada24} for the laboratory
frame). We notice that the directions of $\bm{\Omega}$
and $\bm{J}$ do not coincide in the triaxial body.
The angular momentum vector of the body does not change
in the laboratory frame. The motion of $\bm{\Omega}$ in
the body frame corresponds to the flip motion of the magnetar
body in the laboratory frame, satisfying the angular momentum
conservation.
%
%
\par
If the magnetar's shape does not deform during
the Dhzanibekov effect, the flip happens periodically
\citep{landau69}. The period can be written as
%
\begin{eqnarray}
\tau_{\rm f}\sim200~{\rm yr}~K_{,1}\delta_{,-9}{}^{-1}P_{,0},
\label{eq:t flip}
\end{eqnarray}
%
where $P_{,0}=(P/1~{\rm s})$ is the rotation period of
the second principal axis and $K$ is a numerical dimensionless
factor, respectively.\footnote{
The period depends on the inertial moment and rotation
period of each axis. The numerical factor $K$ summarizes
a combination of these effects.}
The flip takes a time of $\sim0.1\tau_{\rm f}$ in the case
of Figure~\ref{fig:body}. However, this periodicity may
{\it not} be realized in our scenario because of deformations
of the magnetar as expected by \citet{wada24} (i.e., the
rigid body approximation is broken within a finite time).
We suppose that the glitch phenomena are due to the variation
of angular velocity triggered by some non-trivial processes
\citep[e.g.,][]{link07}. The magnetar is not required to
follow fairly the solutions of the Euler equations, which
are used to estimate the breaking of the magnetar's body
in the following.
\par
The deformations of the magnetar mentioned above
may be driven by the inertial force, especially
the Euler force $\bm{F}_{\rm Eul}=\bm{x}\times\dot{\bm{\Omega}}$,
where $\bm{x}$ is the position vector on the magnetar.
The Euler force becomes suddenly stronger in the flip
term of $\Delta\tau_{\rm f}$ than in the interval term
with the period $\tau_{\rm f}$,
according to the variations of $\dot{\bm{\Omega}}$.
The magnitude of the sudden arising Euler force per unit mass
can be estimated by the Euler equations (\ref{eq:euler eq}) as
$F_{\rm Eul}
=|\bm{x}\bm{\times}\dot{\bm{\Omega}}|
\sim R_*\Omega^2\delta$,
where $R_*$ is the radius of the magnetar and
$\Omega=2\pi/P$, respectively. Such a sudden rise of the
Euler force may be large enough to disturb the force
balance between the Lorentz force and elastic force at
the crust, inducing the plastic flow (i.e., non-recovering
deformations in continuum medium are forced). 
\par
The magnitude of $F_{\rm Eul}$ on the body surface
is shown in Figure~\ref{fig:body}, where the value is
normalized by its maximum in the flip term. The Euler force
at a mass element of the body follows the temporal
evolution of  $\bm{\Omega}(t)$ and $\bm{\dot{\Omega}}(t)$
in the body frame.
The directions of the Euler force at a given $t$
are orthogonal to $\bm{\Omega}$ and $\bm{\dot{\Omega}}$
(see the solid and dashed arrows in Figure~\ref{fig:body}),
and orthogonal to the radial direction of the body
($\bm{F}_{\rm Eul}=\bm{x}\times\bm{\dot\Omega}$,
and see also Figure~4 of \citet{wada24}).
The mass elements in the body can be significantly disturbed
by the temporal variations of the Euler force if its magnitude
is sufficiently larger than the rigidity of the body.
\par
It is expected that a part of the crust will be
breaking (or fracturing) when a degree of deformation proceeds some
criteria for the rigidity
\citep[e.g.,][discussed below]{franco00,kojima21}.
We suppose that the plastic flow eventually splits a part of
the crust layer into two or more separated parts. Then, we denote that
``crack" is a material separation made by opening or sliding motions, 
while ``fracture" indicates a region/material significantly affected by
unstable crack growth (or simply breaking part of the crust layer)
in this paper.
\par
At the beginning of the cracks,
the depth of the fractures ${\cal H}$
can be estimated by comparing the total work by the Euler force and
the critical elastic energy. The former can be estimated as
%
\begin{eqnarray}
 W_{\rm Eul}
\sim\frac{\Delta {\cal P}^2}{2M}
\sim\frac{ M R_*{}^2
\Omega^4\delta^2\Delta \tau_{\rm f}{}^2 }{ 2 },
\label{eq:work}   
\end{eqnarray}
%
where $M$ is the mass within a volume of $V\equiv\pi l^2{\cal H}$,
$\Delta{\cal P}=MF_{\rm Eul}\Delta\tau_{\rm f}$
is a momentum gain due to the Euler force,
$l$ is the length scale of the fractures, respectively.
$\Delta\tau_{\rm f}$ is a timescale within which
the rigid body approximation is valid.
From the force balance arguments, \citet{wada24} obtained $l>10^3$~cm.
The latter is
%
\begin{eqnarray}
U_{\rm ela,c}=\mu\sigma_{\rm c}{}^2V,
\label{eq:elastic}
\end{eqnarray}
%
where $\mu$ and $\sigma_{\rm c}$ are the shear modulus and
the critical shear-strain of the crust. We adopt these values
as $\mu/\rho=10^{14}~{\rm cm^2~s^{-2}}(A/56)^{-4/3}(Z/26)^2
\rho_{,4}^{1/3}$~\citep{ogata90,chamel08} and $\sigma_{\rm c}=0.01$.
The critical shear-strain, $\sigma_{\rm c}$, is currently not
fully understood and predicted values range from $10^{-5}$
to $0.1$ depending on approach \citep[summarized in][]{kojima21}.
The density is derived from the hydrostatic equilibrium and
the polytrope equation of state with an index of $1.4$,
%
\begin{eqnarray}
\rho=\rho_{s}
\left(\frac{{\cal H}}{{\cal H}_{s}}\right)^{5/2},
\label{eq:poly}
\end{eqnarray}
%
where we adopt
$\rho_s=10^4$~g~cm$^{-3}$ and ${\cal H}_{s}=10$~cm to reproduce
the density profile derived from effective field theory in nuclear
physics \citep[see][for details]{cehula24}.
Here, we regard that the depth of ${\cal H}$ is equal to the density
scale height $H_\rho\equiv -\rho/(d\rho/dr)$, where $r$
is the radial coordinate of the magnetar.
\par
%
\begin{figure}
\includegraphics[width=\columnwidth]{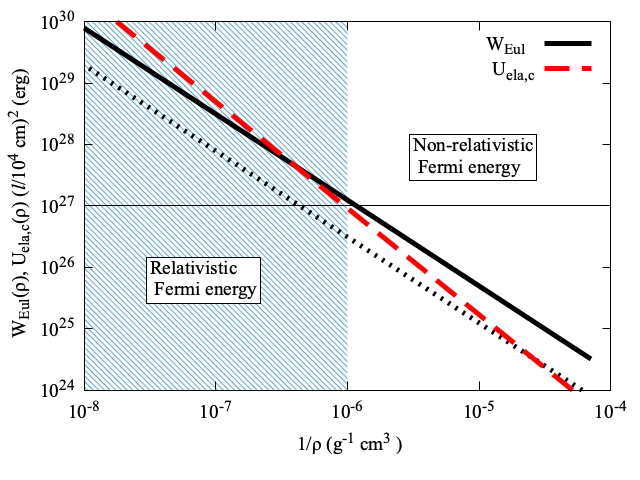}
\caption{
Comparison of the total work of the Euler force $W_{\rm Eul}$
and the critical elastic energy $U_{\rm ela,c}$.
The dots show $W_{\rm Eul}$
at $\Delta\tau_{\rm f}=0.75\times10^{-3}\tau_{\rm f}$,
and the solid line shows one
at $\Delta\tau_{\rm f}=1.5\times10^{-3}\tau_{\rm f}$.
The dashed line shows $U_{\rm ela,c}$.
The hatched region indicates the region where
the Fermi energy of electrons is relativistic
($\rho>10^{6}$~g~cm$^{-3}$).
We adopt
$R_*=10^6$~cm,
$\sigma_{\rm c}=0.01$,
$l=10^4$~cm, and
$M=\rho V$.}
\label{fig:elastic}
\end{figure}
%
Setting $\Delta\tau_{\rm f}\sim1.5\times10^{-3}\tau_{\rm f}$
and $M\sim\rho V$ for simplicity, we find that $W_{\rm Eul}$ is
larger than $U_{\rm ela,c}$  at $\rho\lesssim10^6$~g~cm$^{-3}$ which
corresponds to ${\cal H}\lesssim {\cal H}_{\rm c}=$~63~cm
(Figure~\ref{fig:elastic}).
We find that the Euler force can drive the plastic flow
at the crust, eventually fracturing the crust at
$\rho\lesssim10^6$~g~cm$^{-3}$ within a timescale of
$\Delta\tau_{\rm f}\sim0.3$~yr~$\delta_{-9}P_{,0}$.
The fractures can reach the density scale of $\rho\sim10^6$~g~cm$^{-3}$,
at which a typical energy of the degenerate electrons in the crust
becomes relativistic, implying the leaking of electrons from the cracks
against the gravitational potential.
The estimates for the magnitude of $W_{\rm Eul}$ and $U_{\rm ela,c}$
can be written by introducing $\Delta\tau_{\rm f}=\alpha\tau_{\rm f}$
as, respectively,
%
\begin{eqnarray}
W_{\rm Eul}
&\sim&
6\times10^{30}~{\rm erg}
\nonumber \\
&\times&
\alpha_{,-1}{}^2
\rho_{,6}{}^{7/5}
l_{,4}{}^{2}
R_{*,6}{}^2
P_{,0}{}^{-2}
K_{,1}{}^2,
\label{eq:mag euler}
\end{eqnarray}
%
and
%
\begin{eqnarray}
U_{\rm ela,c}
&\sim&
9\times10^{26}~{\rm erg}
\nonumber \\
&\times&
\rho_{,6}{}^{26/15}
l_{,4}{}^{2}
\sigma_{c,-2}{}^{2}
\left(\frac{Z}{26}\right)^2
\left(\frac{A}{56}\right)^{-4/3}.
\label{eq:mag ela}
\end{eqnarray}
%
Thus, the magnetar is drastically fractured if
the Euler force works full-time of the Dzhanibekov effect.
For $\alpha\sim0.1$, $W_{\rm Eul}=U_{\rm ela,c}$ is given at
unacceptable density scale of $\rho\sim10^{18}$~g~cm$^{-3}$
(or depth of ${\cal H}\sim40$~km), while
the central density of neutron stars is estimated as
$\sim10^{14}$~g~cm$^{-3}$ \citep[e.g.,][]{lattimer01,cehula24}.
The magnetar is no longer regarded
as a rigid body, and the phenomenon may stop
at a time of $\Delta\tau_{\rm f}\ll0.1\tau_{\rm f}$.
\par
The estimated timescale $\Delta\tau_{\rm f}=
1.5\times10^{-3}\tau_{\rm f}$ taken in Figure~\ref{fig:elastic}
is comparable with one of the observed glitches associated
with bursting phenomena
\citep[e.g.,][for PSR J1846-0258]{sathyaprakash24}.
In the case of SGR~1935+2154 with a quicker glitch-burst
association \citep[$\sim10$~hours,][]{hu24}, a larger
ellipticity of $\delta\sim10^{-6}$ is required. 
The X-ray pulse emission analysis implies such a large ellipticity
in magnetars \citep[e.g.,][]{makishima14,makishima24,kolesnikov22}.
It should be noted that although $W_{\rm Eul}$ and $U_{\rm ela,c}$ do
not depend on $\delta$ explicitly, our scenario may be valid at
$\delta\gtrsim10^{-9}$, below which the free precession
approximation is inadequate due to the spin-down of
the magnetar \citep{wada24}.
\par
Here, we summarize our expectations
of the fractures for subsequent bursting phenomena.
We suppose, for simplicity, that the sudden rise of
the Euler force splits a part of the crust into 
separated parts.
As shown in Figure~\ref{fig:elastic},
the magnitude of $W_{\rm Eul}$ increases in time, and
is overcoming 
$U_{\rm ela,c}$
from the outer layer.
Therefore, we expect that
the fracture progresses from the outer layer
to the inner layer in time.
Unless the fracture
reaches at a density scale of 
$\sim10^6~{\rm g~cm^{-3}}$,
above which
the electron kinetic energy (Fermi energy) well exceeds
the gravitational potential energy, the electrons, and ions may
not be ejected from the separated crust. The matter density at
the gap of the separated region (crack) may be significantly small.
When the fracture reaches such a critical density scale of
$\sim10^6~{\rm g~cm^{-3}}$,
the electrons and ions can be expelled into the gap
like a balloon burst, which is the bursting phenomenon studied below.
In the following, we discuss a
further development of such a bursting phenomenon
for observations of photons in section~\ref{sec:burst} 
and 
an ion acceleration at the moment of the beginning of the burst
for the highest-energy CRs in section~\ref{sec:acc}.

\subsection{Available Energy for Magnetar Bursts}
\label{sec:burst}
The internal energy of the crust may be released from the crack
like a balloon burst.
We describe expectations for subsequent
phenomena, focusing on the energetics. In the following, we use
notations of `e' and `i' for electrons and ions
to denote their physical values. For example, the plasma frequency
of a particle symbolically denoted by $o=$\{e,~i\}
is defined as $\omega_{{\rm p},o}=\sqrt{4\pi q_o{}^2 n_o/m_o}$,
where $q_o$, $m_o$, and $n_o$ are the electric charge, rest mass,
and number density of the particle $o$, respectively.
The densities of ions and electrons are calculated as
$n_{\rm i}=\rho/(Am_{\rm p})$ and $n_{\rm e}=Z\rho/(Am_{\rm p})$,
respectively, where $m_{\rm p}$ is the proton rest mass.
The electrons are treated under the cold limit
approximation. The ion temperature is assumed to be
$T_{\rm i}=\hbar\omega_{{\rm p,}i}\simeq0.3$~keV$\rho_{,6}{}^{1/2}$,
where $\hbar$ is the reduced Planck constant,
for simplicity.
\par
Inside the crust, the degenerate electrons can dominate over
the internal energy. The Fermi energy of electrons is
$\epsilon_{\rm F,e}=\sqrt{1+x_{\rm r,e}{}^2}m_{\rm e}c^2$, where
$x_{\rm r,e}=\lambdabar_{\rm e}(3\pi^2n_{\rm e})^{1/3}
\simeq1.0\left(\rho_{,6}Z/A\right)^{1/3}$,
the speed of light is $c$, and  the reduced Compton length
is $\lambdabar_{\rm e}=\hbar/(m_{\rm e}c)$,
respectively~\citep[e.g.,][]{chamel08}.
When the fractures propagate to the 
critical density scale of $\sim 10^6\,{\rm g\,cm^{-3}}$,
that is, the 
depth
of ${\cal H}\sim{\cal H}_{\rm c}=63$~cm$~\rho_{,6}{}^{2/5}$ 
(see equation~\ref{eq:poly}),
the degenerate electrons begin to leak from the bottom of
fractures against gravity.
The available energy for the burst is limited by
the electron internal energy in our scenario.
The internal energy density of the degenerate electrons under the cold
limit approximation, where the chemical potential is equal
to $\epsilon_{\rm F,e}$, is derived from the Fermi distribution
function as
%
\begin{eqnarray}
\varepsilon_{\rm e}
=
\frac{\epsilon_{\rm F,e}}{8\pi^2\lambdabar_{\rm e}{}^3}
\left[
x_{\rm r,e}\left(2x_{\rm r,e}{}^2+1\right)
-\frac{ \ln{\left( x_{\rm r,e}+\sqrt{1+x_{\rm r,e}{}^2}\right) } }{ \sqrt{1+x_{\rm r,e}{}^2} }
\right].
\nonumber \\
\label{eq:interal}
\end{eqnarray}
%
\par
The strong magnetic field of the magnetar, $B\sim10^{15}$~G
is assumed in this paper, restricts the leaking electron's motion 
along the field line unless typical electron energy, $\epsilon_{\rm F,e}$,
exceeds the excitation energy of the first Landau level,
%
\begin{eqnarray}
\epsilon_{\rm L,e}^{(1)}
=m_{\rm e}c^2\left(
\sqrt{1+2\frac{\hbar\omega_{\rm c,e}}{m_{\rm e}c^2}}-1\right)
\simeq5.8m_{\rm e}c^2,
\label{eq:landau}
\end{eqnarray}
%
where $\hbar\omega_{\rm c,e}
=\hbar q_{\rm e}B/m_{\rm e}c\simeq11.6~{\rm MeV}B_{,15}$ is used.
The deexcitations of electrons at higher Landau levels are
triggered as lower levels are released due to the leaking of electrons,
and photons with energy of $\epsilon_{\rm L,e}^{(1)}\simeq5.8m_{\rm e}c^2$
are emitted.
Then, the electron-positron pair creation/annihilation may ignite;
a hot plasma consisting of the electrons, positrons, ions, and photons is
formed at $\rho>\rho_{\rm fb}\sim3.9\times10^{8}$~g~cm$^{-3}$
(${\cal H}>{\cal H}_{\rm fb}=6.9$~m), where $\rho_{\rm fb}$ is derived from
$\epsilon_{\rm F,e}=\epsilon_{\rm L,e}^{(1)}$.
\footnote{If $\epsilon_{\rm L,e}^{(1)}\ll m_{\rm e}c^2$, the
electron and ions may be just thermalized; The internal energy
may not be efficiently released and may be conducted into
the surrounded crust material. The corresponding field strength
is $B\ll10^{14}$~G.}
This generation of the hot plasma corresponds to
releasing the internal energy of electrons at
higher Landau levels.
\par
Such energy release should stop once the number of
excited electrons is small. The decrease in the electron
number in the crust is expected at a density of
$\rho_{\rm ND}\gtrsim4\times10^{11}~{\rm g~cm^{-3}}$
(${\cal H}_{\rm ND}\sim110$~m), where
the electrons interacting with the protons bound in the nuclei
lead to neutron drip off \citep[e.g.,][]{chamel08}.
The electron number density becomes
small and the deexcitations should be suppressed.
Thus, we regard that once the fractures reach 
the density of $\rho\sim\rho_{\rm ND}$,
the new generations of photons due to the deexcitations almost stop.
\par
Then, we consider the effects of the
thermal relaxation of the plasma at
 $\rho_{\rm fb}\lesssim\rho\lesssim\rho_{\rm ND}$
(${\cal H}_{\rm fb}\gtrsim{\cal H}\gtrsim{\cal H}_{\rm ND}$).
The energy density of photons, $\sim T_{\rm ph}{}^4/(\hbar c)^3$,
may increase as the pair creation/annihilation progresses. The cold
ions with the initial internal energy density of
$\sim n_{\rm fb,i}T_{\rm fb,i}$ are heated by the photons, where
$T_{\rm fb,i}=\hbar\omega_{\rm p,i}\simeq6.2$~keV~$\rho_{\rm fb}{}^{1/2}$
and $T_{\rm fb,i}\gg
\left(\epsilon_{\rm F,i}-m_{\rm i}c^2\right)\simeq0.7$~keV
may allow us to approximate the equation of state as the ideal gas.
\footnote{The exact distribution function and relevant equation of state
may deviate from the case of the Maxwell-Boltzmann distribution 
\citep[e.g.,][]{spangler80,gould82}. However, we follow the simplification
used in e.g., \citet{kusunose83}.}
The radiation pressure of the photons affects the ion gas dynamics.
This effect may be significant
when $n_{\rm fb,i}\hbar\omega_{\rm p,i}\sim T_{\rm ph}{}^4/(\hbar c)^3$,
leading to $T_{\rm ph}\sim120~{\rm keV}~\rho_{\rm fb}{}^{3/8}$.
The luminosity of the photons can be $L_{\rm ph}\sim
\pi l^2 cT_{\rm ph}{}^4/(\hbar c)^3
\sim3\times10^{41}$~erg~s$^{-1}l_{,4}{}^2\rho_{\rm fb}{}^{3/2}$.
Since the luminosity is much larger than the Eddington limit for
a solar mass object, $L_{\rm Edd}\sim10^{38}$~erg~s$^{-1}$,
the crust material initially at $\rho\lesssim\rho_{\rm fb}$ may be
expelled.
We note that the energy budget for the photons is the internal energy 
of the electrons at higher Landau levels, not the work done by
the Euler force, which just triggers such bursting activity \citep{wada24}.
\par
%
\begin{figure}
\includegraphics[width=\columnwidth]{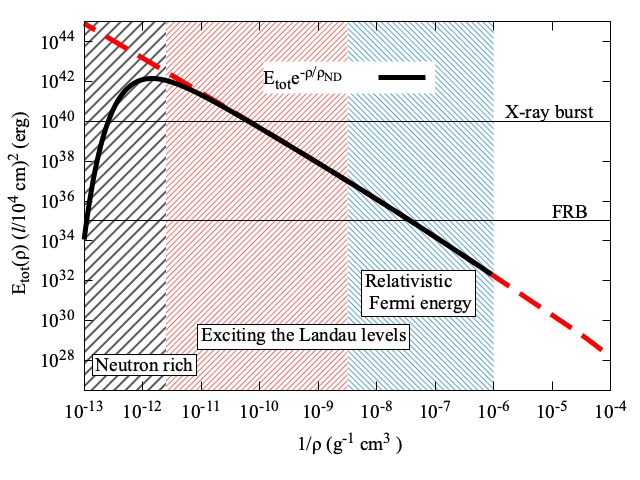}
\caption{
The total internal energy of electrons.
The hatched regions indicate the regions where,
from right to left,
the Fermi energy of electrons is relativistic ($\rho>10^{6}$~g~cm$^{-3}$),
the Fermi energy of electrons is larger than the excitation energy
of the first Landau level ($\rho>10^{8.5}$~g~cm$^{-3}$), and
the neutron drip density $\rho>4\times10^{11}$~g~cm$^{-3}$. 
 We adopt
 $l=10^4$~cm
 and $B=10^{15}$~G, respectively.
 The horizontal lines at $10^{35}$~erg and $10^{40}$~erg
 indicate total energies of the FRB and X-ray short bursts
 for SGR~1935+2154 mentioned in Section~\ref{sec:intro}, respectively.}
\label{fig:energetics}
\end{figure}
%
The available total energy for the burst may be written as
%
\begin{eqnarray}
E_{\rm tot}(\rho)
\sim\pi l^2
\int_{{\cal H}_{\rm s}}^{{\cal H}(\rho)}\varepsilon_{\rm e}d{\cal H}.
\end{eqnarray}
%
Figure~\ref{fig:energetics} shows $E_{\rm tot}(\rho)
{\rm e}^{-\rho/\rho_{\rm ND}}$ for given $l=10^4$~cm and
$B=10^{15}$~G, where we multiply the factor of
${\rm e}^{-\rho/\rho_{\rm ND}}$. The pair plasma creation/annihilation
may stop at this neutron-rich layer with $\rho \gtrsim \rho_{\rm ND}$.
The maximum value is $E_{\rm tot}(\rho_{\rm ND})
=10^{43}$~erg~$l_{,4}{}^2$.
\par
The total energy of the burst may be around
$E_{\rm tot}\sim10^{42}$~erg~$l_{\rm ,4}{}^2$, depending on
the size of cracks, $l$.
The energy release of this bursting phenomenon may take a time of
$\tau_{\rm bst}\sim E_{\rm tot}/L_{\rm ph}\sim10$~s~$\rho_{\rm fb}{}^{-3/2}$,
which does not depend on $l$ but rather depends on $B$.
The maximum Lorentz factor of the outflow, $\Gamma_{\rm bst}$, can be
estimated by equating the photon energy,
$L_{\rm ph}\tau_{\rm bst}\sim E_{\rm tot}$, to the kinetic energy
of the expelled crust, $\sim\left(\Gamma_{\rm bst}-1\right)M_{\rm bst}c^2$.
Then, we obtain $\Gamma_{\rm bst}\sim50$, where
$M_{\rm bst}\sim\rho\pi l^2 {\cal H}$,
the equation~\eqref{eq:poly},
and $\rho\sim\rho_{\rm fb}$ are used.
The estimated total energy and luminosity
are compatible with the FRB-associated burst from
SGR~1935+2154 \citep[e.g.,][]{mereghetti20}.
The maximum Lorentz factor of $\Gamma_{\rm bst}\sim50$
may be preferred to explain the associated
FRB \citep[see e.g.,][for details]{
metzger19,beloborodov20,ioka20,wada23,wada_asano24,iwamoto24}.
\par
To predict the photon spectra and light curves for comparisons of
actual observations, we must simultaneously solve at least
relaxation processes of the highly dense plasma, dynamics of the
expelled crust with the radiation transfer, progressions of the
particle distribution function, and affections
for the inner region of $\rho\gtrsim\rho_{\rm ND}$.
Although we do not perform such analysis in this paper, which seems to be
a tough problem, the observational feature may be speculated as follows.
\par
Here, the outflow is assumed to be observed from its
propagation front for simplicity. The radiation from
the outflow may continue during the energy injection
time of $\tau_{\rm bst}\sim10$~s. The observed photon
energy in this radiation-pressure dominant phase may
be $\sim T_{\rm ph,obs}\sim\Gamma_{\rm bst}T_{\rm ph}
\sim120$~keV.\footnote{
Here, we approximate $T_{\rm ph}{}^4/(\hbar c)^3\gg\rho c^2$ and
$\Gamma_{\rm bst}\gg1$ at the photon escaping time.
Then, supposing a spherical adiabatic expansion with a radius $r$,
the conservation laws of mass flux, momentum flux, and entropy can
be written as
$r^2\rho\Gamma_{\rm bst}={\rm const.}$, $r^2 T_{\rm ph}{}^4
\Gamma_{\rm bst}{}^2={\rm const.}$, and $r^2T_{\rm ph}{}^{3}
\Gamma_{\rm bst}={\rm const.}$, respectively. Then,
we obtain $\Gamma_{\rm bst}\propto r$,  $T_{\rm ph}\propto
r^{-1}$ and $T_{\rm ph,obs}\sim\Gamma_{\rm bst}T_{\rm ph}
\sim{\rm const}$.}
From the relativistic propagation effect, the duration
time or rise time of light curve may be $\tau_{\rm obs}
\sim(c-{\cal V}_{\rm bst})\tau_{\rm bst}/c\sim\tau_{\rm bst}/\Gamma_{\rm bst}{}^2
\sim2$~ms~$(\tau_{\rm bst}/10~{\rm s})(\Gamma_{\rm bst}/50)^{-2}$,
where ${\cal V}_{\rm bst}/c\equiv\sqrt{1-1/\Gamma_{\rm bst}{}^2}
\simeq1-1/2\Gamma_{\rm bst}{}^2$ is used.
Since there is a finite time for the acceleration of outflow
up to $\Gamma_{\rm bst}\sim50$, the duration time may be
$\tau_{\rm obs}>2$~ms in reality. Thus, the light curve may be
observed as a short pulse with a duration time/rise time of $\gtrsim2$~ms,
and typical photon energy may be $\sim120$~keV. These can be compatible
with peaks seen in giant flares
\citep[e.g.,][]{hurley98,hurley05,mereghetti24}
and the FRB-associated burst from
SGR~1935+2154 \citep{mereghetti20}.
The above estimates are independent of the total energy,
$E_{\rm tot}\propto l^2$, which can be as large as
$\sim10^{46}$~erg~$(l/10^6~{\rm cm})^2$ in our model.
Then, as suggested by \citet{thompson95}, the outflow may disturb
the magnetosphere of the magnetar, leading to the repeating pulse
emissions as seen in the flares.
\par
\citet{ofek08} pointed out a possible miss-identification of
a giant flare from a nearby galaxy as the gamma-ray burst GRB~070201
\citep[see also,][]{abbott08}. It implies that giant flares resemble
GRBs in their light curves.
The maximum available energy of
$E_{\rm tot}\lesssim10^{46}$~erg~$(l/10^{6}~{\rm cm})^2$
seems too tiny to be observed as GRBs, though comparable
with the giant flares.
Thus, our scenario does not predict an intrinsic relationship between
giant flares and GRBs unless an additional energy source becomes available.
\par
The deposited energies in tangled magnetic fields
can be such an additional energy source \citep[e.g.,][]{perna11}.
The fields can be formed around the crust layer, as implied by
one of the latest simulations of core-collapse supernovae \citep{nakamura24}.
If the sudden activation of
the crust affects the inner core region of $\rho\gtrsim\rho_{\rm ND}$,
the tangled fields can `push' the crust with their magnetic pressure.
The deposited energy in the tangled fields should be limited by the
gravitational potential energy of
$U_{\rm grav}\sim GM_*{}^2/R_*\sim5\times10^{53}$~erg.
Thus, if an energy of $\sim0.01$-$0.1$~$U_{\rm grav}$ becomes available,
the total energy of the burst can reach one estimated in GRBs.
\par
We currently can {\it not} conclude whether the bursting
phenomena studied here correspond to known/already-observed transient
events. Further studies along the lines mentioned above should
be performed in the future.

\section{Acceleration of Ions}
\label{sec:acc}
As mentioned in previous studies \citep[e.g.,][]{arons03,asano06,kotera11},
magnetar's outflow (wind) is expected as one of the sources of
ultrahigh-energy CRs. The problem pointed out by the authors
themselves is that there is no firm explanation of the ion injection
process in the acceleration region. In our scenario, the ion injection
is naturally expected. If the rotation period of the magnetar is
$P\sim 1$~ms as the authors assumed
(e.g., a newborn neutron star is considered), some ions
can be accelerated in the magnetospheric
voltage drops across the magnetic field \citep{arons03,kotera11}
or in the shock waves \citep[][the so-called internal shock model
is adopted]{asano06}.
\par
We suggest another idea of the ion acceleration mechanism 
associated with the magnetar burst in our scenario, which does
not assume such a small rotation period of $P\sim 1$~ms.
Here, we consider a specific situation: the crack
appears near the magnetic pole and lies perpendicularly to
the magnetic field line.
\par
\begin{figure}
\includegraphics[width=\columnwidth]{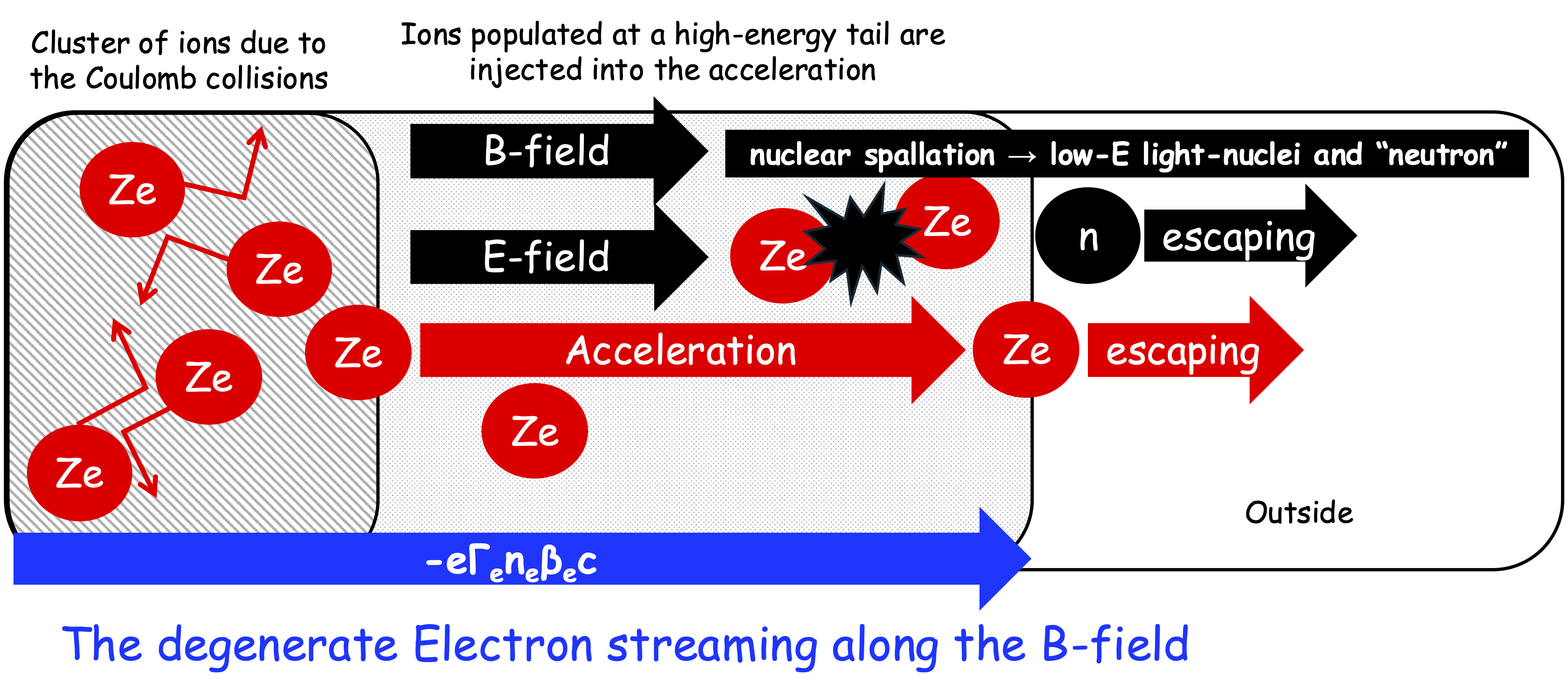}
\caption{Schematic illustration of the ion acceleration
due to the self-discharge effects by the degenerate
electron streaming. The electric field accelerates
the ions and may be immediately screened out by
the daughter particles of nuclear spallation with
a time scale of $\sim1$~ps.}
\label{fig:schematic}
\end{figure}

Initially, the degenerate electrons 
stream along the magnetic field
at $\rho<\rho_{\rm fb}$, while the ions are not relativistic
and not degenerate; $x_{\rm r,i}=(m_{\rm e}/m_{\rm i})Z^{1/3}
x_{\rm r,e}\simeq2.9\times10^{-5}\rho_{,6}{}^{1/3}$,
$(\epsilon_{\rm F,i}-m_{\rm i}c^2)\simeq0.02$~keV, and
$T_{\rm i}=\hbar\omega_{\rm p,i}\simeq0.3$~keV$\rho_{,6}{}^{1/2}$.
The idea of ion acceleration is similar to the effect of
`self-discharge' by streaming high-energy particles
\citep[e.g.,][]{ohira22}. Figure~\ref{fig:schematic}
shows a schematic illustration of the effect.
%
%
\par
We consider a moment at the beginning of the burst phenomenon
when the Euler force makes the fracture at the depth of
${\cal H}\sim{\cal H}_{\rm c}=63$~cm ($\rho\sim\rho_{\rm c}
=10^{6}$~g~cm$^{-3}$). The initial number flux of the leaking
electrons may be $\Gamma_{\rm e}n_{\rm e}\beta_{\rm e}c\sim
cx_{\rm r,c} n _{\rm e,c}$
(i.e., $\Gamma_{\rm e}=\sqrt{1+x_{\rm r,c}{}^2}$),
where $\beta_{\rm e}\simeq 1$ is the electron linear velocity
divided by the speed of light, $\Gamma_{\rm e}=
1/\sqrt{1-\beta_{\rm e}{}^2}$,
$n_{\rm e,c}=Z\rho_{\rm ,c}/Am_{\rm p}$, and
$x_{\rm r,c}=x_{\rm r,e}(\rho_{\rm c})$, respectively.
\par
The electric current density due to the electrons is
$j_{\rm e}=-e\Gamma_{\rm e}n_{\rm e}\beta_{\rm e}c$, where $e$
is the elementary charge, $n_{\rm e}$ is
the electron number density at the rest frame of electrons.
Below, we use such a conventional form \citep[e.g.,][]{BM76}:
number density, temperature, and particle energy are measured
in the rest frame, while the others
are measured in the laboratory frame.
Note that the electric current densities
of both electrons and ions are measured in the laboratory
frame in this expression.
The ions should compensate for
the charge separation, so their current becomes
$j_{\rm ret}=Ze\Gamma_{\rm i}n_{\rm i}\beta_{\rm i}c=-j_{\rm e}$.
However, the compensation is delayed due to the ion-ion collisions.
Since the state space of ions is sparsely occupied
($T_{\rm i}\gg\epsilon_{\rm F,i}$), the ion-ion collisions dissipate
the ions' coherent motion along the magnetic field. As a result, the
electric field is induced to compensate for the charge separation
and the field strength can be estimated by
equating the acceleration term due to the electric field and
the dissipation term due to the collision,
$\Gamma_{\rm i}n_{\rm i}q_{\rm i}E\sim
-m_{\rm i}\Gamma_{\rm i}n_{\rm i}\beta_{\rm i}c\nu_{\rm C,ii}$ as
(Ohm's law)
%
\begin{eqnarray}
E_{\rm res}
\sim
-\frac{m_{\rm i}}{q_{\rm i}}\beta_{\rm i}c\nu_{\rm C,ii}
\sim
\frac{ m_{\rm i}                 }{ Z^2e                    }
\frac{ \nu_{\rm C,ii}         }{ \Gamma_{\rm i}n_{\rm i} }
cx_{\rm r,e}n_{\rm e,c},
\label{eq:res}
\end{eqnarray}
%
where we use $j_{\rm ret}=-j_{\rm e}$
and suppose the conservation of the number flux,
$\Gamma_{\rm e}n_{\rm e}\beta_{\rm e}c\sim x_{\rm r,e}n_{\rm e}c$,
and
%
\begin{eqnarray}
\nu_{\rm C,ii}
\sim
4\pi \Gamma_{\rm i}n_{\rm i}c
\left(
\frac{  Z^2e^2 }{ m_{\rm i}c^2 }
\right)^2
\left(
\frac{ T_{\rm i} }{ m_{\rm i}c^2 }
\right)^{-3/2},
\label{eq:ii}
\end{eqnarray}
%
is the ion-ion Coulomb collision rate. Here, we omit the
Coulomb logarithm for simplicity. This estimate may be valid
for the ions running around the `head' of current.
The field strength of
$ZeE_{\rm res}\sim5.6\times10^{22}$~eV~cm$^{-1}$
$\rho_{\rm c}{}^{4/3}$
$(Z/26)^{13/3}$
$(A/56)^{-5/6}$
$(T_{\rm i}/0.3~{\rm keV})^{-3/2}$
is much smaller than the critical strength
given by the classical radiation loss limit\footnote{
It should be noted that the radiation reaction or the self-force problem
may still be open questions
\citep[e.g.,][]{hammond10PRA,cole18,ares18,adam21}.}
of $ZeE_{\rm rad}\sim 3q_{\rm i}{}^2/2r_{\rm cl,i}{}^2
\sim43\times10^{24}$~eV~cm$^{-1}(Z/26)^{-2}(A/56)^2$,
where $r_{\rm cl,i}\equiv q_{\rm i}{}^2/m_{\rm i}c^2$.
\par
We suppose that the electric field estimated above
maintains only a very short timescale as $\sim1$~ps (derived later).
So, only a very small fraction of the ions may be accelerated
at the moment of the begining of the burst.
The electron stream can also be decelerated
by the electric field in the case of collisionless plasma,
where the charged-particles motions are determined by only
the electromagnetic fields.
However, in our case, the electrons in the stream occupy
the momentum space significantly (an extreme limit
of the dense, collisional plasma).
The electrons coming from the inner
part may occupy a larger momentum-space volume than
those ejected in a prior time.
The electrons may not return to the original location, and
the charge separation may be sustained unless the ions
compensate for it. It is expected that, however,
the nuclear spallation of the accelerated ions
injects daughter ions, electrons, and positrons.
The electric field may be immediately screened out by
the daughters. Hence, we expect that the fraction
and maximum energy of the accelerated ions
may be limited by the nuclear spallation time scale.
\par
The ions with an energy of $\epsilon_{\rm i}\gg T_{\rm i}$,
populating at a high-energy part of their Maxwell-Boltzmann
distribution, can run away from
the cluster of the colliding ions and be injected into the
acceleration. We parameterize the injection energy by $\eta>1$
as $\epsilon_{\rm inj}=\eta T_{\rm i}$,
giving the number density of ions into the acceleration as
$n_{\rm inj}\sim n_{\rm i}{\rm e}^{-\eta}$.
Note that the $\eta$ may be a function of time in reality.
The acceleration should be stopped when the ions undergo
the nuclear spallation reaction. We simply set the cross-section
as $\sigma_A\sim\pi a_{A}{}^2$, where
$a_{A}=4.59\times10^{-13}~{\rm cm}(A/56)^{1/3}$ is the radius of
stable nucleus \citep[e.g.,][]{krane87}, and  the mean free path
is $l_{\rm mfp}\sim1/n_{\rm inj}\sigma_A$.
Thus, the maximum energy of the accelerating ion can reach
$\sim10^{21}~{\rm eV}={\rm ZeV}$ scale;
%
\begin{eqnarray}
\epsilon_{\rm i,max}
&\sim& 
ZeE_{\rm res}l_{\rm mfp}
\\
&\sim&
1.2~{\rm ZeV}
\nonumber \\
&\times&
\rho_{\rm c}{}^{1/3}
\left(\frac{Z}{26}\right)^{13/3}
\left(\frac{A}{56}\right)^{-1/2}
\left(\frac{T_{\rm i}}{0.3~{\rm keV}}\right)^{-3/2},
\nonumber
\end{eqnarray}
%
where we suppose $\eta=5$ (${\rm e}^{\eta}\simeq150$).
The acceleration takes a time of
$\sim l_{\rm mfp}/c\sim7.0\times10^{-13}~{\rm s}=0.7~{\rm ps}$
($l_{\rm mfp}\simeq0.02$~cm).
\par
Even if the fraction of ions in the acceleration is small ($\eta\gg1$)
and the electric field becomes strong as $E_{\rm res}\sim E_{\rm rad}$
due to non-trivial effects (e.g., detailed progressions of the leaking
plasma, probabilistic experience of the nuclear spallation reactions,
and so on), the maximum energy may be limited by the depth of fractures
${\cal H}_{\rm c}\sim63$~cm; the pair plasma at the magnetosphere
may screen out the electric field. Thus, the maximum energy can {\it not}
be larger than $ZeE_{\rm res}{\cal H}_{\rm c}
<ZeE_{\rm rad}{\cal H}_{\rm c}
\sim2.7\times10^{27}$~eV, near the Planck energy scale
of $1.2\times10^{28}$~eV. The acceleration takes a time of
$\sim {\cal H}_{\rm c}/c\sim2~$ns in this case.
\par
When the fraction of the daughters due to
the nuclear spallation increases ($t\gtrsim1$~ps),
the electric field may be screened out. The maximum energy of
the ions may immediately decrease. The acceleration of ions
by the self-discharge effect may also terminate.
\par
The energy spectral distributions of the nuclei
and consequent emissions are particularly interesting in the context of
cosmic-ray origin and astrophysical transient events.
Although detailed processes (such as the acceleration, the escape of nuclei
from the fractures, and the effects of the spallation reactions)
are not studied here, we estimate the spectrum shape and
the upper limit of total energy very roughly below.
\par
The spectral distribution of the accelerating nuclei in a uniform
electric field is proportional to $\epsilon_{\rm i}{}^{-1}$
\citep[e.g.,][]{arons03}.
If we set the injection  number density to be
$n_{\rm inj}=\rho{\rm e}^{-\eta}/m_{\rm i}$
with fixed $\eta$ as one of the most idealized cases,
we may obtain
${\rm d}N_{\rm i}/{\rm d}\epsilon_{\rm i}{\rm d}\rho
=(\rho{\rm e}^{-\eta}/m_{\rm i})\epsilon_{\rm i}{}^{-1}
\exp\left[-\epsilon_{\rm i}/\epsilon_{\rm i, max}(\rho,\eta)\right]
\delta_D(\rho-\rho_{\rm c})$, where $\delta_D(\rho)$ is
Dirac's delta function.
Then, the spectrum is
%
\begin{eqnarray}
\frac{{\rm d}N_{\rm i}}{{\rm d}\epsilon_{\rm i}}
=
\frac{\rho_{\rm c}{\rm e}^{-\eta}}{m_{\rm i}}
\epsilon_{\rm i}{}^{-1}
\exp\left[
   -\frac{ \epsilon_{\rm i} }{ \epsilon_{\rm i, max}(\rho_{\rm c},\eta) }
   \right],
\end{eqnarray}
%
where $\eta$ and $\epsilon_{\rm i, max}$
may be functions of time in reality.
The total energy of the accelerated ions may
be up to (taking fixed $\eta=5$)
%
\begin{eqnarray}
E_{\rm acc,ion}
&\ll& V
\int \epsilon_{\rm i}
\frac{{\rm d}N_{\rm i}}{{\rm d}\epsilon_{\rm i}} {\rm d}\epsilon_{\rm i}
\\
&\simeq&
7.0\times10^{41}~{\rm erg}
\nonumber \\
&\times&
\rho_{\rm c}{}^{1/3}l_{,4}{}^2
\left(\frac{Z}{26}\right)^{13/3}
\left(\frac{A}{56}\right)^{1/2}
\left(\frac{T_{\rm i}}{0.3~{\rm keV}}\right)^{-3/2},
\nonumber
\end{eqnarray}
%
where $V=\pi l^2l_{\rm mfp}$ is used
for the upper limit estimate (the gap of the crack may have an
area much smaller than $\pi l^2$ ).
Thus, up to $\sim10$~\% of the total available energy
$E_{\rm tot}(\rho_{\rm ND})\simeq10^{43}$~erg can
be consumed for the ion acceleration.
Note that in a realistic situation,
$\eta$ and $\epsilon_{\rm i,max}$
may vary to suppress the total energy significantly.
The spectra of consequent radiations
from the accelerating nuclei (photons, neutrinos, neutrons, etc.) 
may show a typical energy scale reflecting the peak energy
of $\epsilon_{\rm i}{}^2{\rm d}N_{\rm i}/{\rm d}\epsilon_{\rm i}$, although
the net spectrum of nuclei may be drastically changed due to the nuclear
spallation reactions. The shape of the spectra may provide hints
of the processes during the acceleration.
\par
The phenomena discussed above involve many physical processes
diversely, and we may still not catch all of the related or
consequent phenomena and processes.
The FRBs observed in SGR~1935+2154 might be related to
the processes at the beginning of the burst in terms of energetics (Figure~\ref{fig:energetics}).
Moreover, the secondary nuclei are emerging from nuclear
spallation reactions, whose temporal evolution is not studied here.
The analysis of our model is also related to strong-field
quantum electrodynamics seen in high-intensity lasers,
particle/nuclear physics at parameter spaces still not
sufficiently explored, and condensed matter physics in
extreme situations, similar to the bursting phenomenon
discussed in Section~\ref{sec:burst}.
The model should be studied along such lines in the future.

\section{Prospects for Observations}
\label{sec:discussion}
To discuss our model separately from the others, we name
the phenomena along with our scenario as
`Instant ZeV-ion-acceleration in Upset Magnetar Origin bursts'
(IZUMO bursts). Further analysis of a burst-to-burst rate
density of short bursts/giant flares and further surveys of
hard X-ray/soft gamma-ray transients by such as
{\it the Einstein Probe} mission \citep{yuan22} and
{\it FORCE} mission \citep{force22} are required to constrain
the IZUMO burst by observing photons.
\par
The cross-matching studies of the ultrahigh-energy CR observations
and magnetar bursts can be valuable in finding the IZUMO bursts.
When the IZUMO burst is located at external galaxies, we have
too many candidates within a backtracked localization of
ultrahigh-energy CR, as shown by \citet{unger24} for the case
of Amaterasu \citep[see also,][]{zhang_theodore24,bourriche24}.
In the case of the IZUMO burst taking place in the Milky Way Galaxy
and/or its satellites, the arrival directions of the ultrahigh-energy
CRs can coincide with the location of magnetars.
The expected gyro-radius of ultrahigh-energy CR is
$r_{\rm g}\sim100~{\rm kpc}~Z^{-1}(\epsilon_{\rm i}/100~{\rm EeV})
(B_{\rm ISM}/1~{\rm \mu G})^{-1}$, which can be much larger than the
Galactic radius of $\sim10$~kpc depending on $Z$. Since the nuclear
spallation reactions limit the maximum energy of accelerated ions in
the IZUMO burst, high-energy lighter nuclei and `neutrons' are also
expected. If the case,
some high-energy CR light-nucleus/neutron events with an energy
of $\lesssim(\epsilon_{\rm i,max}/A)
\sim20$~EeV show good coincidence with the IZUMO bursts
in arrival directions and time.
The neutrino events like KM3-230213A with an energy of
$\sim0.05\epsilon_{\rm i,max}/A\sim100$~PeV
due to the nuclear spallation reactions (at the source or
during the propagation) may also be expected.
The CR experiments, such as the Telescope Array \citep{abu-zayyad12,tokuno12}
and the Pierre Auger Observatory \citep{auger15}
can constrain the details of the IZUMO bursts.
Note that the arrival time and direction coincidence/correlation among
the $\sim20$~EeV CR neutrons,
$\sim100$~PeV neutrinos, and $\sim100$~keV X-rays
from neutron stars in nearby galaxies is the essential prediction of our scenario at the present
rather than detailed observed flux and spectral shape of them.
Future observations will constrain the scenario.
\par
\citet{TA_prl24,TA_prd24} reported one of the latest
results of the arrival direction and compositions of ultrahigh-energy CRs
with $>100$~EeV and concluded the composition at $>100$~EeV is very heavy.
\citet{zhang_theodore24} and \citet{bourriche24} also pointed out that
if Amaterasu is a heavy nucleus such as iron, its source can be distributed
far from the local void. The IZUMO bursts can be consistent with these recent
results and implications.
\par
We consider a required source density in the Universe
so that the IZUMO burst becomes a non-negligible source of the observed
CRs above $100$~EeV. From the observed CR spectrum around the Earth,
the source density for CRs above $100$~EeV is 
${\cal N}\gtrsim2\times10^{-5}~{\rm Mpc^{-3}}$
\citep{TA_prd24}. Considering the magnetar birth rate density of
$\sim10^{-6}$ Mpc$^{-3}$ yr$^{-1}$ \citep[e.g.,][]{globus23} and
its characteristic age of $\tau_{\rm ch}\sim 10$~kyr
\citep[e.g.,][]{olausen14}, the number density of magnetar can be
$\sim10^{-2}$~Mpc$^{-3}$. Thus, $\gtrsim0.2$~\% of the magnetars
should be an effective host of the IZUMO burst. The fraction of
$\gtrsim0.2$~\% may be realized since
$\tau_{\rm f}/\tau_{\rm ch}\sim 0.2\delta_{,-9}{}^{-1}P_{,0}$,
see, Equation~\eqref{eq:t flip}. Note that \cite{ligo22} gave an upper limit
of the ellipticity as $\delta\lesssim10^{-6}$ from observations of
continuous gravitational waves. 

\section{Conclusions}
\label{sec:conclusions}
We have studied the 
possibilities of highest-energy CR
acceleration in a novel scenario for the bursting activity
of magnetar,
in which the magnetar undergoes the Dzhanibekov effect.
This effect can result in a sudden rise of the Euler force,
and a part of its crust is assumed to be fractured. Then,
the magnetar's internal energy can be released like a balloon burst.
The total energy, light curve, and typical photon energy of the bursts
can be compatible with the observed bursts.
The ions can be accelerated at the beginning of the burst
and the maximum energy can reach $\sim$ZeV.
The scenario includes several assumptions and simplifications,
but it can be tested by the observations of $\sim20$~EeV CR neutrons,
$\sim100$~PeV neutrinos, and $\sim100$~keV photons.
Further studies covering broad regions of physics are required.



\begin{acknowledgments}
We are grateful to K. Asano, T. Sako, and the anonymous referee
for their comments that improve the paper.
We also thank T. Kawashima for the daily discussion.
This work is supported by the joint research program of the Institute for
Cosmic Ray Research (ICRR), the University of Tokyo, and KAKENHI grant
Nos. 24K00677 (J.S.) and 22K20366 (T.W.).
\end{acknowledgments}

%

\vspace{5mm}
\facilities{}


\software{}         



\appendix


\bibliography{papers}{}
\bibliographystyle{aasjournal}



\end{document}